\documentclass{ws-procs9x6}

\def\met{\mathbin{E\mkern - 11mu/_T}}

\begin{document}

\title{Top Physics Results at CDF}

\author{TREVOR VICKEY\footnote{\uppercase{O}n behalf of the \uppercase{CDF} \uppercase{C}ollaboration.}}

\address{University of Illinois at Urbana-Champaign \\
Department of Physics \\
1110 West Green Street \\ 
Urbana, IL 61801-3080, USA\\ 
E-mail: Trevor.Vickey@cern.ch}

\maketitle

\abstracts{The most recent results on top quark physics at CDF are reported.  Measurements of cross-section and mass are presented, and the status of single top quark production searches are discussed.  The results obtained from probing various top quark properties are also presented.}

\section{Introduction}
Recently discovered\cite{top_disc} by the CDF and D0 Collaborations, the top quark is the least well understood fundamental particle.  As a consequence of its enormous mass, around 175 GeV$/$c$^2$, this hefty particle has a lifetime shorter than the hadronization timescale.

With the top mass close to the scale of electroweak symmetry breaking there are hints of an intimate relationship between top and this mechanism--Higgs is most strongly coupled to the top quark.  By studying top we are testing electroweak theory, as this quark may be sensitive to physics beyond the standard model (SM).

\section{Top Quark Production and Decay Modes}
At the Tevatron, a $p\overline{p}$ collider with a center of mass energy of $\sqrt{s} = 1.96$ TeV, top is predominantly pair produced via $q\overline{q}\rightarrow t\overline{t}$ 85\% of the time with the remaining fraction generated via $gg\rightarrow t\overline{t}$.  The cross-section for $t\overline{t}$ production at the Tevatron, assuming $m_t = 175$ GeV$/$c$^2$, is\cite{top_theory} $6.7^{+0.7}_{-0.9}$ pb.  By comparison, the total cross-section for producing top singly via electroweak processes is smaller by about a factor of two.

Top decays to a $W$ boson and a $b$ quark nearly 100\% of the time due to unitarity constraints on the CKM matrix.  The experimental signature of a $t\overline{t}$ event includes two $b$ quark jets, the presence of multiple light-quark jets and/or a single high-$p_T$ charged lepton accompanied by significant missing transverse energy ($\met$) from an undetected neutrino.  Top candidate events are frequently classified by the $W$ boson decays.  The ``dilepton'' mode occurs when each of the two $W$s decay leptonically.  Both $W$s decaying hadronically produce the ``all-hadronic'' mode; events with a mixture of leptonic and hadronic decays are dubbed ``lepton-plus-jets'' events.

\begin{figure}[t]
\centerline{\epsfxsize=3.5in\epsfbox{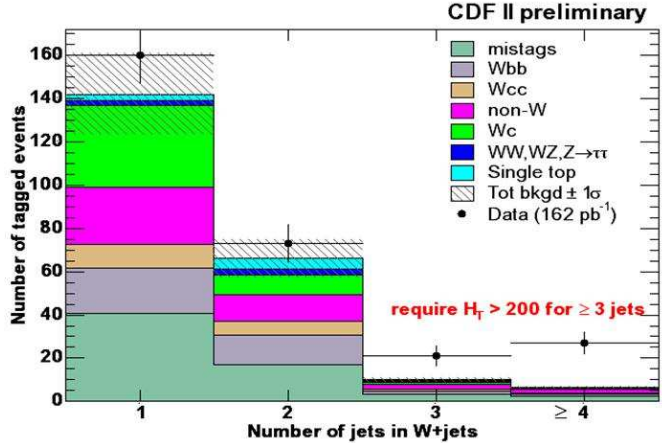}}   
\caption{The top quark signal region of the lepton-plus-jets sample occupies the 3 and $\geq 4$ jet bins.\label{fig:num_of_jets}}
\end{figure}

\section{Top Quark Cross-section Measurements}

\begin{table}[hb]
\tbl{Top quark pair-production cross-section measurements.}
{\footnotesize
\begin{tabular}{@{}llc@{}}
\hline
{} &{} &{}\\[-1.5ex]
{} & \multicolumn{1}{c}{$\sigma_{t\overline{t}}$ [pb]} & $\int\mathcal{L}\mbox{ }dt$ [pb$^{-1}$] \\[1ex]
\hline
{} &{} &{}\\[-1.5ex]
Dilepton: Combined                    & $7.0^{+2.7}_{-2.3}\mbox{ (stat.) }^{+1.5}_{-1.4}\mbox{ (syst.)}$ & 200\\[1ex]
Dilepton: $\met$, Num. jets           & $8.6^{+2.5}_{-2.4}\mbox{ (stat.) }\pm 1.1\mbox{ (syst.)}$ & 200\\[1ex]
\hline
{} &{} &{}\\[-1.5ex]
Lepton + Jets: Kinematic              & $4.7\pm 1.6\mbox{ (stat.) }\pm 1.8\mbox{ (syst.)}$ & 193\\[1ex]
Lepton + Jets: Kinematic NN           & $6.7\pm 1.1\mbox{ (stat.) }\pm 1.6\mbox{ (syst.)}$ & 193\\[1ex]
Lepton + Jets: Vertex Tag + Kinematic & $6.0\pm 1.6\mbox{ (stat.) }\pm 1.2\mbox{ (syst.)}$ & 162\\[1ex]
Lepton + Jets: Vertex Tag             & $5.6^{+1.2}_{-1.1}\mbox{ (stat.) }^{+0.9}_{-0.6}\mbox{ (syst.)}$ & 162\\[1ex]
Lepton + Jets: Double Vertex Tag      & $5.0^{+2.4}_{-1.9}\mbox{ (stat.) }^{+1.1}_{-0.8}\mbox{ (syst.)}$ & 162\\[1ex]
Lepton + Jets: Jet Probability Tag    & $5.8^{+1.3}_{-1.2}\mbox{ (stat.) }\pm 1.3\mbox{ (syst.)}$ & 162\\[1ex]
Lepton + Jets: Soft Muon Tag          & $5.2^{+2.9}_{-1.9}\mbox{ (stat.) }^{+1.3}_{-1.0}\mbox{ (syst.)}$ & 193\\[1ex]
\hline
{} &{} &{}\\[-1.5ex]
All Hadronic: Vertex Tag              & $7.8\pm 2.5\mbox{ (stat.) }^{+4.7}_{-2.3}\mbox{ (syst.)}$ & 165\\[1ex]
\hline
\end{tabular}\label{tab:xsec_table} }
\vspace*{-13pt}
\end{table}

Conducting a top quark production cross-section measurement validates our top-enriched samples and could yield the first signs of new physics in the top quark sector.  

The dilepton analyses use event selections of either two identified charged leptons ($e$ or $\mu$), at least two jets and large $\met$ or a single charged lepton in addition to a well isolated track.  

Analyses in the lepton-plus-jets channel have a larger initial data sample than dilepton analyses, albeit one which contains a larger amount of background contamination.  Tagging $b$-jets is a technique used to increase the fraction of top in the lepton-plus-jets sample.  The tagger used by CDF is sensitive to displaced secondary vertices due to the relatively long-lived $b$ quarks, in addition to semi-leptonic decays of the $b$.  The signal region for tagged lepton-plus-jets events is shown in Fig.~\ref{fig:num_of_jets}. 

The $b$ tagging is also used by the all-hadronic analysis in addition to a trigger which requires four high $p_T$ jets and a large total transverse energy in the event.

Results of CDF's most recent cross-section measurements are summarized in Table~\ref{tab:xsec_table}.  All results obtained thus far are consistent with the SM.\cite{cross_section}

\section{Top Quark Mass Measurements}
Through the correlation with other SM parameters, a measurement of the top mass puts constraints on the Higgs.  A precise top mass measurement is difficult due to the level of understanding necessary concerning jet energies and the relationship between parton-level objects and detector observables.  This measurement is also hampered by our ability to correctly assign jets to the parton-level objects in top decays.

\begin{table}[ht]
\tbl{Top quark mass measurements.}
{\footnotesize
\begin{tabular}{@{}llc@{}}
\hline
{} &{} &{}\\[-1.5ex]
{} & \multicolumn{1}{c}{$m_t$ [GeV$/$c$^2$]} & $\int\mathcal{L}\mbox{ }dt$ [pb$^{-1}$] \\[1ex]
\hline
{} &{} &{}\\[-1.5ex]
Dilepton: $\phi$ of $\nu$       & $170.0\pm 16.6\mbox{ (stat.) }\pm 7.4\mbox{ (syst.)}$ & 193\\[1ex]
Dilepton: $p_z$ $t\overline{t}$ & $176.5^{+17.2}_{-16.0}\mbox{ (stat.) }\pm 6.9\mbox{ (syst.)}$ & 193\\[1ex] 
Dilepton: $\nu$ weighting       & $168.1^{+11.0}_{-9.8}\mbox{ (stat.) }\pm 8.6\mbox{ (syst.)}$ & 200\\[1ex]
\hline
{} &{} &{}\\[-1.5ex]
Lepton + Jets: Multivariate     & $179.6^{+6.4}_{-6.3}\mbox{ (stat.) }\pm 6.8\mbox{ (syst.)}$ & 162\\[1ex]
Lepton + Jets: $M_{\rm reco}$   & $177.2^{+4.9}_{-4.7}\mbox{ (stat.) }\pm 6.6\mbox{ (syst.)}$ & 162\\[1ex]
Lepton + Jets: DLM              & $177.8^{+4.5}_{-5.0}\mbox{ (stat.) }\pm 6.2\mbox{ (syst.)}$ & 162\\[1ex]
\hline
\end{tabular}\label{tab:mass_table} }
\vspace*{-13pt}
\end{table}

Several methods are used by CDF to measure the top mass in the dilepton and lepton-plus-jets channels.  Most of the analysis techniques use templates generated from Monte Carlo events in conjunction with likelihood fitting.  The Dynamical Likelihood Method (DLM), however, takes into account all possible jet combinations in an event and the likelihood is multiplied event-by-event to derive the top quark mass using a maximum likelihood method.  The advantage of the DLM method over the canonical template methods lies in the fact that the cross-section is used as a posterior probability whereas in the template methods it is used as a prior probability.  

Results of the most recent top mass measurements are summarized in Table~\ref{tab:mass_table}.

\section{Single Top Quark Searches}
Investigating single top production is a great opportunity to study the charged-current weak interaction and to search for new physics thought to be exclusive to these channels.  Single top production cross-sections are proportional to the CKM matrix element $V_{tb}$; a measurement of the cross-section provides a direct measurement of this quantity.  CDF has searched for the $s$ and $t$-channel single top production modes which have theoretical cross-sections of 0.88 and 1.98 pb, respectively.\cite{single_top_theory}

The strategy for single top analyses is to search for $W$ decay products plus two or three jets.  Two analyses were conducted in 162 pb$^{-1}$ of data to search for single top.  A combined search using the scalar sum of the event transverse energy ($H_T$ distribution) was used for single top discovery and to measure $|V_{tb}|$.  Separate $s$ and $t$-channel searches using the $Q\times\eta$ distribution\footnote{$Q$ is the charge of the lepton and $\eta$ of the light-quark jet.} were carried out to reveal  any new physics.  

The combined search sets a limit of $< 17.8\mbox { pb @ } 95\%\mbox{ CL}$.  Limits from the $s$ and $t$-channel searches are $< 13.6\mbox { pb @ } 95\%\mbox{ CL}$ and $<10.1\mbox { pb @ } 95\%\mbox{ CL}$, respectively.\cite{single_top}

\section{Measurements of Top Quark Properties}

Now that sizable samples of top quark candidate events have been accumulated, we can proceed to measure various top quark properties.

The fraction of right-handed $W$ bosons from top decay is heavily suppressed in the SM.   The charged-lepton $p_T$ and angular distributions for each of the three $W$ helicity states are very distinct--a feature exploited to make a helicity measurement using templates in likelihood fits to the CDF data.  CDF uses both the charged-lepton $p_T$ and lepton angular distributions for extracting the fraction of longitudinal $W$s.  The lepton $p_T$ analysis carries out a measurement in both the dilepton and lepton-plus-jets datasets measuring $F_0 = 0.27^{+0.35}_{-0.21}\mbox{ (stat.) }\pm0.17\mbox{ (syst.)}$, the angular distribution method uses the lepton-plus-jets dataset measuring $F_0 = 0.89^{+0.30}_{-0.34}\mbox{ (stat.) }\pm 0.17\mbox{ (syst.)}$.  CDF has recently revisited the Run I data to make a measurement of the right-handed fraction,\cite{w_helicity} $F_+ < 0.18\mbox{ @ }95\%\mbox{ CL}$.

An analysis of top decay kinematics\cite{anom_kin} yields results consistent with the SM.  CDF has measured several ratios of branching fractions:  $BR(t\rightarrow\tau\nu b)/BR_{\rm SM}(t\rightarrow\tau\nu b) < 5.0\mbox{ @ }95\%\mbox{ CL}$ and $BR(t\rightarrow WB)/BR(t\rightarrow Wq) > 0.62\mbox{ @ }95\%\mbox{ CL}$.  We have also measured the ratio of the cross-sections $\sigma_{\rm dilepton}/\sigma_{\rm lepton + jets} = 1.45^{+0.83}_{-0.55}\mbox{(stat. + syst.)}$.

\section{Outlook}
Experimentally, top quark physics is still in its infancy.  While no unexpected results have been observed thus far, many opportunities for discovery still exist at CDF. 

As CDF continues the trend of doubling its dataset each year, statistical and systematic uncertainties will be reduced greatly improving the current measurements.

\end{document}